\input harvmac.tex

\def\p{\partial}
\def\a{\alpha}
\def\b{\beta}
\def\g{\gamma}
\def\d{\delta}

\def\th{\theta}
\def\n{{n}}
\def\x{\xi}
\def\t{\tau}
\def\z{\zeta }
\def\D{\Delta }
 
\def\[{\left[}
\def\]{\right]}
\def\({\left(}
\def\){\right)}
\def\<{\left\langle\,}
\def\>{\,\right\rangle}
 \def\hf{ {\frac{1}{2}}}
\def\half{ {1\over 2} }
 \def\frac#1#2{ {{\textstyle{#1\over#2}}}}
\def\IR{{ \Bbb R} }

  \def\tz{\tilde z}

\def\gst{\gamma _{\rm str}}

\def\DB{ \Delta^{_B} }
\def\ss{{\vec S}} 
\def\AA{D^{\perp}_0}\def\BB{D^{||}_1}
%

  \lref\KazPerc{ V.~K.~Kazakov, ``Percolation on a fractal with the
  statistics of planar Feynman graphs: Exact solution'', Mod.\ Phys.\
  Lett.\ A {\bf 4}, 1691 (1989).  }
  \lref\KazIz{ V.~A.~Kazakov, ``Exact solution of the Ising model on a
  random two-dimensional lattice'', JETP Lett.\ {\bf 44}, 133 (1986)
  [Pisma Zh.\ Eksp.\ Teor.\ Fiz.\ {\bf 44}, 105 (1986) ].  }
  \lref\BK{ D.~V.~Boulatov and V.~A.~Kazakov, ``The Ising Model On
  Random Planar Lattice: The Structure Of Phase Transition And The
  Exact Critical Exponents,'' Phys.\ Lett.\ {\bf 186B}, 379 (1987).  }
 \lref\DK{B. Duplantier and I. Kostov, {Phys.  Rev.  Lett} {\bf 61}
 (1988), 1433; { Nucl.  Phys.} {\bf B 340} (1990) 491.  }
\lref\Ion{ I.~K.~Kostov, ``O(n) Vector Model On A Planar Random
Lattice: Spectrum Of Anomalous Dimensions,'' Mod.\ Phys.\ Lett.\ A
{\bf 4}, 217 (1989).  }
\lref\ADEold{ I.~K.~Kostov, ``The $ADE$ face models on a fluctuating
planar lattice" Nucl.\ Phys.\ B {\bf 326}, 583 (1989).  }
\lref\DTS{ I.~K.~Kostov, ``Strings with discrete target space,''
Nucl.\ Phys.\ B {\bf 376}, 539 (1992) [arXiv:hep-th/9112059].  }
 \lref\PolyakovL{A. Polyakov, ``Quantum Geometry Of Bosonic Strings",
 {Physics Letters} {\bf B103 } (1981), 207.  }
\lref\KPZ{ V.~Knizhnik, A.~Polyakov and A.~Zamolodchikov, {Mod.  Phys.
Lett.} {\bf A3}, 819 (1988); F.~David, {Mod.  Phys.  Lett.} {\bf A3},
1651 (1988); J.~Distler and H.~Kawai, {Nucl.  Phys.} {\bf B321}, 509
(1989).  }
\lref\DDK{ F.~David, ``Conformal field theories coupled to 2-D gravity
in the conformal gauge'', { Mod.\ Phys.\ Lett.}\ {\bf A 3}, 1651
(1988);\ J.~Distler and H.~Kawai, ``Conformal field theory and 2-D
quantum gravity or who's afraid of Joseph Liouville?'', {Nucl.\
Phys.}\ {\bf B 321}, 509 (1989).  }
\lref\CardyBB{ J.~L.~Cardy, ``Conformal Invariance And Surface
Critical Behavior,'' { Nucl.\ Phys.}\ {\bf B 240}, 514 (1984).  }
 \lref\CardyOn{ J.~L.~Cardy, ``The O(n) model on the annulus,''
 [arXiv:math-ph/0604043].  } \lref\Duplantierev{ B.~Duplantier,
 ``Conformal fractal geometry and boundary quantum gravity,''
 [arXiv:math-ph/0303034].  }
  \lref\nienhuis{B. Nienhuis, Phys.  Rev.  Lett.  {\bf 49}, 1062
  (1982); {J. Stat.  Phys.  } {\bf 34}, 731 (1984).  }
 \lref\SLE{See, for example, G.F.~Lawler, {\sl Conformally Invariant
 Processes in the Plane}, Mathematical Surveys and Monographs, Volume 114 (American Math.  Soc., April 2005).  }
    \lref\FZZb{V.~Fateev, A.~B.~Zamolodchikov and A.~B.~Zamolodchikov,
    ``Boundary Liouville field theory.  I: Boundary state and boundary
    two-point function'', [arXiv:hep-th/0001012].  }
\lref\PTtwo{B.~Ponsot, J.~Teschner, ``Boundary Liouville field theory:
Boundary three point function'', Nucl.~Phys.~{\bf B 622} (2002) 309,
[arXiv:hep-th/0110244].  }

\lref\hosomichi{K.~Hosomichi, "Bulk-Boundary Propagator in Liouville
Theory on a Disc", JHEP {\bf 0111}, 044 (2001),
[arXiv:hep-th/0108093].  }
\lref\KPS{ I.~K.~Kostov, B.~Ponsot and D.~Serban, ``Boundary Liouville
theory and 2D quantum gravity'', {Nucl.\ Phys.}\ {\bf B 683}, 309
(2004), [arXiv:hep-th/0307189].  }
 \lref\JS{J.~Jacobsen and H.~Saleur, ``Conformal boundary loop
 models'', [arXiv:math-ph/0611078].}
 %
 \lref\DupSal{B.~Duplantier and H.~Saleur, {Phys.  Rev.  Lett.} {\bf
 58}, 2325 (1987).  }
 \lref\SalBau{H.~Saleur and M.~Bauer, {\it Nucl.  Phys.} {\bf B 320},
 591 (1989).  }
\lref\KKopen{ V. Kazakov and I. Kostov , ``Loop gas model for Open
Strings", {Nucl.Phys.} {\bf B 386}, 520 (1992).}
  \lref\Kbliou{ I.~K.~Kostov, ``Boundary correlators in 2D quantum
  gravity: Liouville versus discrete approach,'' Nucl.\ Phys.\ B {\bf
  658}, 397 (2003) [arXiv:hep-th/0212194].  }
    \lref\Nichols{A. Nichols, V. Rittenberg and J. de Gier, J. Stat.
    Mech.  {\bf 0503}, P003 (2005), [arXiv:cond-mat/0411512]; A.
    Nichols, J. Stat.  Mech.  {\bf 0601}, P003 (2006), hep-th/0509069;
    A. Nichols, J. Stat.  Mech.  {\bf 0602}, L004 (2006),
    [arXiv:hep-th/0512273].  }
%
%
\lref\Pearce{ P.~A.~Pearce, J.~Rasmussen and J.~B.~Zuber,
``Logarithmic minimal models,'' J.\ Stat.\ Mech.\ {\bf 0611}, P017
(2006) [arXiv:hep-th/0607232].  }
\lref\AS{A. Y. Alekseev and V. Schomerus, Phys.  Rev.  D {\bf 60},
061901 (1999); [arXiv:hep-th/9812193].  }
  \lref\Idense{I. Kostov, ``Strings embedded in Dynkin diagrams'',
  Lecture given at Cargese Mtg.  on Random Surfaces, Quantum Gravity
  and Strings, Cargese, France, May 27 - Jun 2, 1990, In *Cargese
  1990, Proceedings, Random surfaces and quantum gravity* 135-149;
  preprint SACLAY-SPH-T-90-133.  }
\lref\EK{ B.~Eynard and C.~Kristjansen, ``Exact solution of the O(n)
model on a random lattice,'' {\it Nucl.\ Phys.} {\bf B 455}, 577
(1995), [arXiv:hep-th/9506193].  }
 \lref\bershkut{ M. Bershadsky and D. Kutasov, ``Scattering of open
 and closed strings in (1+1)-dimensions", {\it Nucl.\ Phys.}\ {\bf B
 382},{213}, ( 1992), hep-th/9204049.  }
%
\lref\BGR{ I.~K.~Kostov, ``Boundary ground ring in 2D string theory'',
{ Nucl.\ Phys.}\ {\bf B 689}, 3 (2004), [arXiv:hep-th/0312301].  }
\lref\KoZa{I. Kostov and Al.  Zamolodchikov, work in progress.}
%
%

\input amssym.def \input amssym.tex 
\input epsf

\font\ninerm=cmr9
\font\ninei=cmmi9

\font\sixi=cmmi6

  \hfuzz 15pt
 \overfullrule=0pt
\Title{\vbox{\baselineskip12pt\hbox
{SPhT-T07/036}\hbox{hep-th/0703221}}}
{\vbox{\centerline
 { Boundary  Loop Models }
\centerline{ and 2D Quantum Gravity}
 \vskip2pt
}}
  \centerline{
 Ivan Kostov\footnote{$^\star$}{Associate member of the Institute for
 Nuclear Research and Nuclear Energy, Sofia, Bulgaria
 }
}

 \vskip 1cm

\centerline{\it Service de Physique Th{\'e}orique, CNRS -- URA 2306,}
\vskip -2pt

\centerline{\it C.E.A. - Saclay,
  F-91191 Gif-Sur-Yvette, France
 }


\vskip 1.5cm \parskip=2pt \baselineskip=13pt 
{ We study the $O(\n)$
loop model on a dynamically triangulated disk, with a new type of
boundary conditions, discovered recently by Jacobsen and Saleur.  The
partition function of the model is that of a gas of self and mutually
avoiding loops covering the disk.  The Jacobsen-Saleur (JS) boundary
condition prescribes that the loops that do not touch the boundary
have fugacity $\n\in [-2,2] $, while the loops touching at least once
the boundary are given different fugacity $y$.  The class of JS
boundary conditions, labeled by the real number $y$, contains the
Neumann ($y=\n$) and Dirichlet ($ y=1$) boundary conditions as
particular cases.  Here we consider the dense phase of the loop gas,
where we compute the two-point boundary correlators of the $L$-leg
operators with mixed Neumann-JS boundary condition.  The result
coincides with the boundary two-point function in Liouville theory,
derived by Fateev, Zamolodchikov and Zamolodchikov.   The Liouville 
charge of the boundary operators match, by the KPZ correspondence, 
with the L-leg boundary exponents conjectured by JS.}

\Date{}
  
\vfill
\eject

 \baselineskip=14pt plus 1pt minus 1pt

 \newsec{Introduction}
 
\noindent The solvable statistical models that have a geometrical
description in terms of self and mutually avoiding clusters, like
Ising, $O(\n)$ and Potts models, can be also formulated and solved on
a dynamical lattice \KazPerc\KazIz \BK\DK \Ion\ADEold\DTS. A
statistical model defined on a dynamical lattice is said to be coupled
to gravity, since the sum over lattices gives a discretization of the
path integral over Riemann metrics on the world sheet.  For each
critical point described by a `matter' CFT, the `coupling to gravity'
consists in adding a Liouville and ghost sectors and dressing the
scaling operators by exponents of the Liouville field
\PolyakovL\KPZ\DDK. The description of the critical points based on
Liouville theory allows to interpret the wealth of exact results about
statistical models on dynamical lattices obtained via matrix-model or
combinatorial techniques.
 
   A statistical system on a random lattice exhibits qualitatively the
same critical phases as that on a regular lattice, but with different
critical exponents.  At a critical point characterized by a conformal
anomaly $c\le 1$, the conformal weight $h$ of a matter field\foot{Only
operators without spin ($h =\bar h $) survive after coupling to
gravity.} can be extracted from its `gravitational dimension' $\D$,
which determines the scaling properties of the correlation functions
involving this field \KPZ\DDK,
\eqn\KPZsc{h  = {\Delta(\Delta-\gst)\over 1-\gst}, \qquad c =1-6 {\gst
^{2} \over 1-\gst}.  } The exponent $\gst$ (gamma-string) describes
the critical fluctuations of the area of the random surface. 
 
The solutions of the KPZ scaling relation \KPZsc\ can be parametrized
by a pair of numbers $r $ and $s$, not necessarily integers:
\eqn\scaldims{ \eqalign{ h  _{rs}=\bar h  _{rs} = {({r /b}- s
b)^2 - ({1/ b } - b)^2 \over 4 }, \qquad \D_{rs} = {{r /b} -
sb -({1/ b}-b)\over 2b}\, .  } } 
The dependence on the matter central charge is through the positive
parameter $b<1$, defined as
 \eqn\defb{ {b^2}\equiv {1\over 1
-\gst}<1 .  } 
The conformal weights have the symmetry $h_{rs}= h_{-r,-s} $, unlike
the gravitational dimensions.  To each conformal weight one can
associate two gravitational dimensions, $\D_{rs} $ and $\D_{-r, -s}$,
which are the two roots of the quadratic relation \KPZsc.  They
correspond to the two possible gravitational dressings of the matter
conformal field by Liouville vertex operators.

The correspondence between the critical phenomena on flat and
dynamical lattices is particularly useful in presence of {boundaries}.
The boundary critical exponents \CardyBB\ on flat and dynamical
lattices are again related by \KPZsc.  Boundary 2D quantum gravity
became a powerful tool for evaluating exact critical exponents
\Duplantierev, complementary to the Coulomb gas techniques \nienhuis\
and the Schramm-Loewner Evolution (SLE) \SLE. For systems coupled to
gravity it is possible, by cutting open the path integral along
cluster boundaries, to solve analytically problems which whose exact
solution is inaccessible on flat lattice.  On the other hand, our
understanding of the boundary phenomena in 2D gravity is much helped
by the progress achieved in boundary Liouville theory in the last
years \FZZb\PTtwo\hosomichi.

In this paper we will focus on the possible boundary conditions and
the spectrum of boundary exponents of the $O(\n)$ model coupled to 2D
gravity \Ion.  Our main purpose is to check a very interesting
conjecture made in \JS\ about the general boundary conditions for the
$O(n)$ model on a flat lattice.

The $O(\n)$ model has a continuum transition if the number of flavors
is in the interval $-2\le \n\le 2$, where it can be parametrized by an
angle,
  \eqn\defnu{ \n=2\cos(\pi\theta ).  }
The boundary $O(\n)$ model was originally considered with {\it Neumann
boundary condition} (the loops are reflected from the boundary).  The
boundary scaling dimensions of the $L$-leg operators, realized as
sources of $L$ open lines, were conjectured in \DupSal\ and then
derived in \SalBau.  Furthermore, the partition function of the
$O(\n)$ model on the annulus with Neumann boundary conditions was
evaluated in \CardyOn.  Another obvious boundary condition is the {\it
Dirichlet boundary condition}, studied in \KKopen\KPS, for which there
is an open line ending at each site of the boundary.\foot{In these
papers the loop gas was considered in the context of the SOS model,
for which the Dirichlet and Neumann boundary conditions have the
opposite meaning.} The dimensions of the $L$-leg boundary operators
with mixed Dirichlet and Neumann boundary conditions were computed in
\Kbliou\KPS\ by coupling the model to 2D gravity and then using the
KPZ scaling relation \KPZsc.
  
Recently, Jacobsen and Saleur put forward a proposal about the
complete classification of the boundary conditions of the $O(\n)$ loop
gas model in the dense phase, described by, in general non-rational,
CFT with $b^2 = 1-\th$.  Their proposal is based on a previous work
\Nichols\ and possibly overlaps, for rational $\th$, with the results
of \Pearce.  The claim of \JS\ is that there is a continuum of
boundary conditions characterized by a real variable $y$.  The
Jacobsen-Saleur boundary condition, which we will denote in the
following by JS, is defined by counting the loops that touch at least
once the boundary with different fugacity $y$, while the loops that do
not touch the boundary are counted with fugacity $\n $.  The JS
boundary conditions contain as particular cases Neumann ($y=\n$) and
Dirichlet ($y=1$) boundary conditions for the $O(\n)$ field.

In order to evaluate the $L$-leg exponents, the authors of \JS\
considered the loop gas on an annulus, with Neumann boundary condition
on the inner rim, JS boundary condition on the outer rim, and $L$
non-contractible loops separating the two boundaries.  The loop model
with these boundary conditions were called there boundary loop model
(BLM).  According to \JS, BLM with $L$ non-contractible lines has two
sectors, the {\it blobbed} and the {\it unblobbed} one, characterized
by two different for ($L\ge 1$) critical exponents.  In the blobbed
(unblobbed) sector the outmost non-contractible line touches at least
once (does not touch) the outer rim.  It is argued in \JS\ that the
scaling exponents in the two sectors are characterized by the
conformal weights
  \eqn\starde{\eqalign{ h  _L^{^ {\rm unblob}}&= h  _{-r, -r+L}\cr h 
  _L^{^ {\rm blob}}\ \ &= h  _{r, r+L} } \qquad ({\rm dense \ phase}),
  }
where the real parameter $r$ is related to the boundary fugacity $y$
by\foot{Our notations are related to those used \JS\ by $\th= {\g\over
\pi} = {1\over p+1}$.}
   \eqn\defy{
 y(r)  = {\sin[(r+1)\pi \theta  ]\over \sin ( r\pi \theta )} .
 }
The boundary exponents for Neumann-Neumann \DupSal\ and
Neumann-Dirichlet \KPS\ boundary conditions appear as particular cases
of \starde\ when $r=1$ and $r= {1-\theta \over 2\theta } $,
correspondingly.

Inspired by \JS, in this paper we analyze the JS boundary conditions
for the $O(\n)$ loop gas coupled to 2D gravity.  Put in string theory
terms, the problem we address is to describe the ensemble of D-branes
in bosonic string theory whose target space is the $(n-1)$-dimensional
sphere.  We restrict ourselves to the dense phase of the $O(\n)$
model, where we derive loop equations in the form of recurrence
relations between the the boundary two-point functions of the $L$-leg
and $(L-1)$-leg operators.  The loop equations are obtained by cutting
open the world sheet along segments of loops that connect two points
of the boundary.  In the continuum limit, the solution of the loop
equations is given, up to a normalization, by the two-point boundary
correlator in boundary Liouville theory \FZZb.  The scaling exponents
characterizing the solution reproduce, {\it via} KPZ relation \KPZsc ,
the spectrum of boundary conformal weights conjectured by JS \JS.

The paper is organized as follows.  In sect.  2 we give a short review
of the results we will need about the loop gas coupled to gravity.  In
particular, we derive the loop equations for the disk partition
function with Neumann boundary conditions.  In sec.  3 we formulate
the JS boundary conditions in terms of the boundary measure for the
$O(\n)$ spins and derive the loop equations for the two-point
correlators with mixed Neumann-JS boundary conditions.  In section 4
we take the continuum limit of the loop equations and evaluate the
boundary $L$-leg exponents.  Here we also show that the loop equations
can be cast in the form of the functional equations, generated by the
boundary ground ring of Liouville gravity.  Summary of the results and
some concluding remarks are presented in Section 5.

 \newsec{The boundary $O(n)$ loop model coupled to 2D gravity }

 \subsec{Disk partition function with Neumann boundary conditions}
 
\noindent The $O(\n)$ model \nienhuis, defined originally on a regular
hexagonal lattice, can be considered on an arbitrary trivalent planar
graph $\CT^*$, which is dual to a triangulation of the disk $\CT$.
The local fluctuating field is a $\n$-component spin $\ss(r)$
associated with the vertices $r$ of $\CT^*$.  The Boltzmann weight for
given configuration is
\eqn\actOn{
\prod_{<\!  rr'\! > } \(t+  
 \ss(r) \cdot \ss(r')\),
}
where the product is over the links $<\!\!  rr'\!\!>$ of $\CT^*$.
Here we consider the maximally packed dense phase $t=0$.  The
partition function for given triangulation is the trace over these
weights, defined by $\Tr [ 1] =1,\ \Tr [ S_a (r ) S_b (r) ]= \d_{ab}$
and $\Tr [S_a(r)] = \Tr [S_a(r) S_b(r)S_c(r)]=0$.  Expanding the trace
as a sum of monomials, the trace for given triangulation $\CT$ can be
written as a sum over all configurations of densely packed
self-avoiding, mutually avoiding loops on $\CT^*$ (Fig.1).  Each loop
is taken with a weight $\n$.

 
 \vskip 20pt
\centerline{  \epsfxsize=100pt \epsfbox{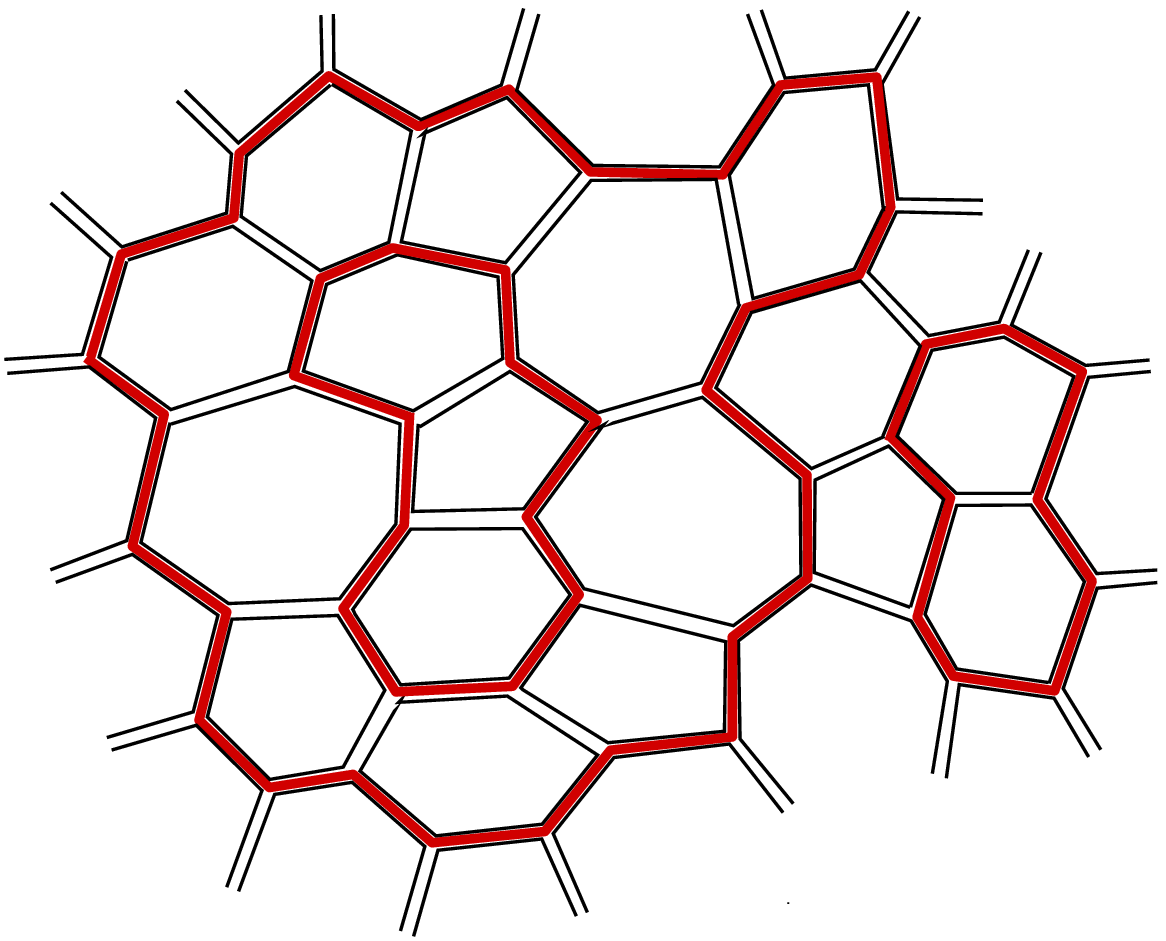}\hskip 1cm 
  \epsfxsize=100pt
 \epsfbox{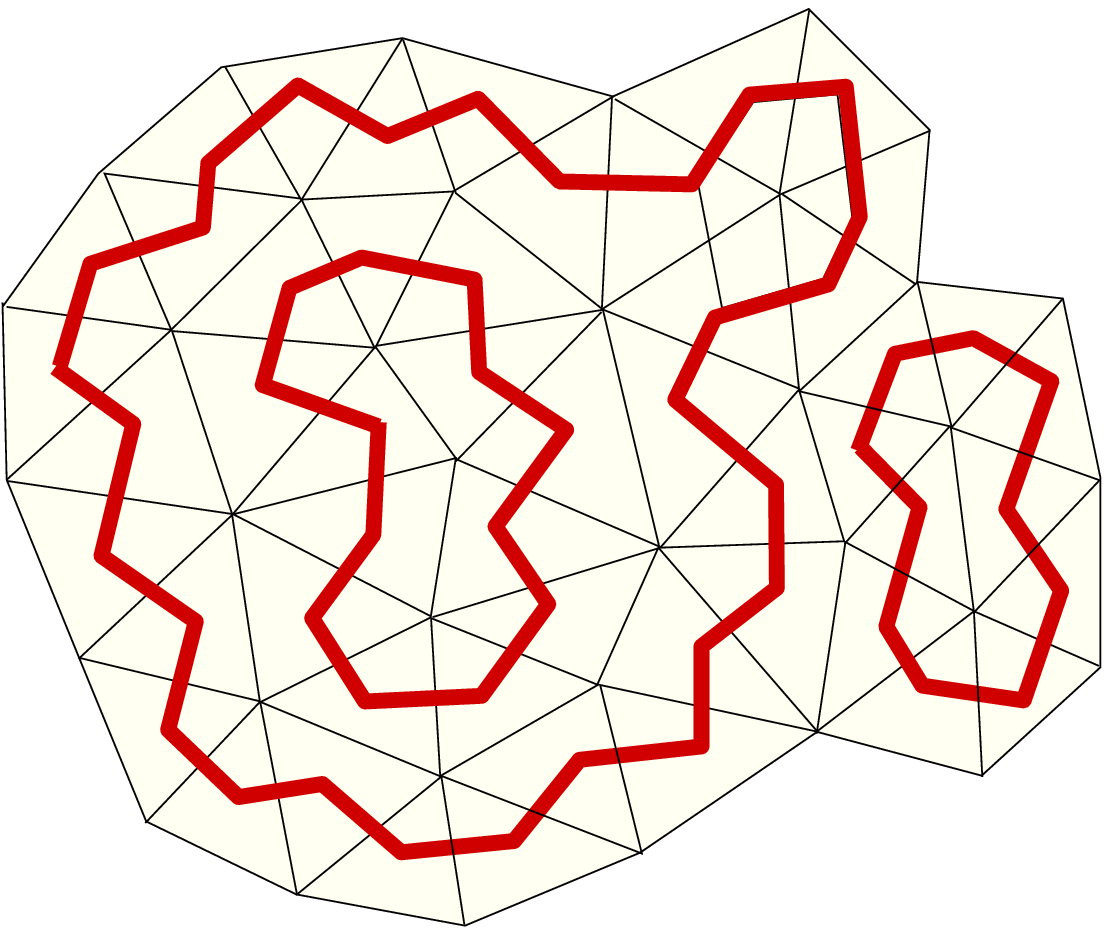}    }
 \vskip 5pt
 
  {\ninepoint Fig. 1:   Loops on   
  a  trivalent planar graph  $\CT^*$ (left)   dual to 
  a triangulation $\CT$ (right)
  \par\hskip 1.3cm
  }
  
  \vskip  10pt
 
The disk partition function of the $O(\n)$ model coupled to 2D gravity
is defined as the double average: with respect to the $O(\n)$ field on
given triangulation and with respect a sufficiently large class of
triangulations.  We will consider all possible triangulations,
including the degenerate ones, which are dual to trivalent planar
graphs with several connected components.  In such a triangulation a
boundary edge can either belong to a triangle, or be identified with
another edge of the boundary.  The measure in the ensemble of
triangulations of the disk is determined by the `cosmological
constant' $x$ coupled to the area and the `boundary cosmological
constant' $\zeta$ coupled to the boundary length.  By definition the
area of a triangulation is the number of its triangles and the
boundary length is the number of edges along the boundary.
   
The basic observable in the $O(\n)$ model coupled to 2D gravity is the
disk partition function $\Phi(\zeta, x)$.  Its derivative $W=-\p_\z
\Phi $ is given by the series
    \eqn\defWz{ W(\z) = \sum_{k=0}^\infty \zeta^{-l} \ W_l(x), }
where $W_l$ is the non-normalized expectation value of the Boltzmann
factor \actOn\ in the ensemble of triangulations $\CT_l$ with boundary
of length $l$.  Since there is no restriction for the $O(\n)$ spins at
the boundary, we have Neumann boundary condition.  The loop expansion
of $W_l$ is
 \eqn\lpgz{ W_l= \sum_{\CT_l} x^{-{\rm (Area \ of } \ \CT_l)} \sum
 _{{\rm loops\ on\ }\CT_l^*} \, \n^{(\rm {Number \ of \ loops})}\, .
 }
Note that the here boundary has a marked point, hence there is no
symmetry factor $1/l$ in the sum.  The generating function \defWz\ is
the boundary one-point function of the identity operator.

 \subsec{Loop equation for the  disk partition function}

\noindent Here we remind the combinatorial derivation of the loop
equations for the the disk amplitude with Neumann boundary conditions
\defWz .  This derivation, first given in \ADEold, is a useful
exercise to do before passing to the more complicated case of mixed
boundary conditions.
  
The triangulations filled by loops that enter in the sum in the r.h.s.
of \lpgz\ can be divided into two classes (Fig.  2).  The first class
comprises the degenerate triangulations for which first boundary edge
is not an edge of a triangle, but is connected directly to the
$k+1$-th boundary edge.  The contribution of such triangulations
factorizes to $W_{k-1} W_{l-k-1}$.  There are $l-1$ such terms, with
$k=1, 2,..., l-1$.  For the rest of the triangulations entering the
sum in \lpgz\ the first edge belong to a triangle, which must be is
visited by a loop, since the loops are densely packed.  Now consider
the ensemble of all triangles visited by this loop.  These triangles
form a closed strip that contains the loop.  Let $q$ and $p$ be
respectively the lengths of the internal and external boundaries of
the strip.  For given $q$ and $p$ the expectation value factorized to
the contribution of the internal disk $W_q $, that of the external
disk $W_{l-1-p}$, and the number of realizations of the strip, $
{(p+q)!\over p!q!}$.  We have also a factor $\n$ because of the loop
contained in the strip.  Adding the contributions of the two classes
of triangulations, we obtain the following bilinear equation for the
$W_k$,
\eqn\pictWw{ \eqalign{
 &W_l = \sum _{k=1}^{l-1} W_{k-1}W_{l-k-1} +\n \sum_{p,q\ge 0}
 {(p+q)!\over p!q!} (1/x)^{p+q+1} W_{l+p-1}W_q , } }
where by definition $W_0=1$.  Equations of this type are usually
called loop equations.  in terms of the generating function \defWz,
the loop equation reads
\eqn\lopeqa{ \[ -\zeta W(\zeta) + W (\zeta)^2+ \n W
(x-\zeta)W(\zeta)\]_{<}=0, } 
where $[\ \ ]_<$ denotes the negative part
of the Laurent series in $\zeta$.

 \vskip  20pt

 \epsfxsize=320pt 
\vskip 20pt
\hskip 10pt
\epsfbox{  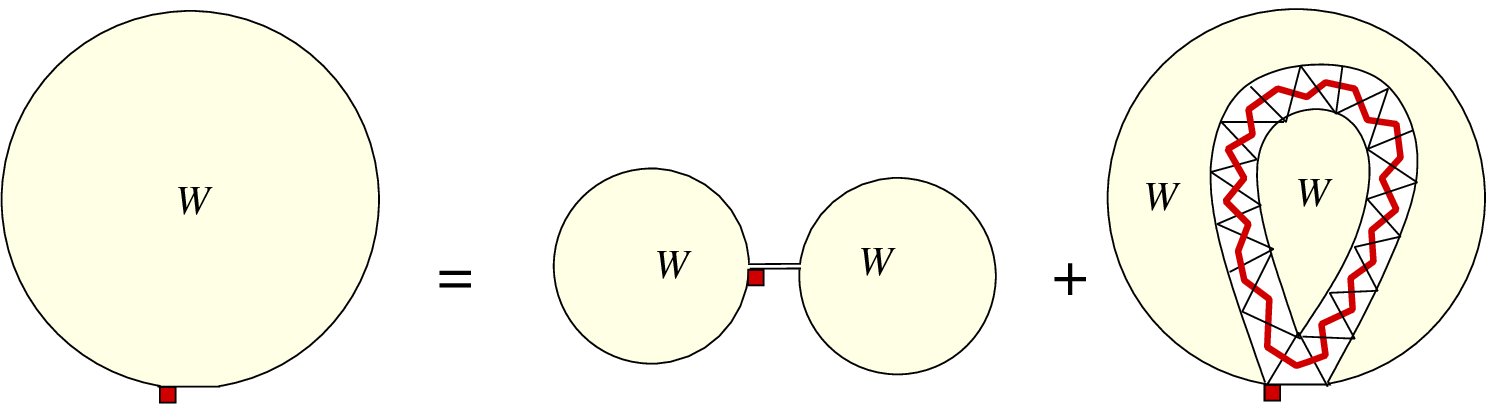   }
\vskip 5pt

\centerline{\ninepoint Fig.  2: Loop equation for the disk partition
function with Neumann boundary  }
\centerline{\ninepoint  \ \ conditions. The small rectangle 
 stands for  the marked point on the boundary. }
 \vskip  30pt
 
  \noindent

\noindent 
It is consistent to assume that the solution has a cut
$[a,b]$ on the real axis, with $a<b< x/2$.  Then one can write the
projection $[\ \ ]_<$ in \lopeqa\ as a contour integral,
\eqn\lopeqW{ 1-\z W(\z) + W^2(\z) +  \n\, \oint {d\z'\over 2\pi i}\, 
{W(\z)-W(\z')\over z-z'} W(x-\z')=0.  } 
where the contour of integration encircles the cut of $W(\z)$ and
leaves outside the cut of $W(x-\z)$. 
Equation \lopeqW\  yields a condition
for the discontinuity across the real axis
\eqn\cutD{ {\rm Disc} W(\z)\, [ -\z + W(\z+i0) +W(\z-i0)+ \n \
W(x-\z)]=0 , \ \ (\z\in \IR).} 
The condition \cutD, after being symmetrized with respect to $\z\to
x-\z$, implies that certain bilinear combination of $W(\z)$ and
$W(x-\z)$ has zero discontinuity on the real axis and therefore is
analytic in the whole complex plane.  This leads, taking into account
the asymptotics $W(\z)\simeq 1/\z$ at infinity, to the functional
identity \DTS
\eqn\wwt{ W(\z)^{2}+W(x-\z)^{2} + \n W(\z)W(x-\z) = \z W(\z) +(x-\z)
W(x-\z)-2.}

\newsec{ Boundary  correlators with  mixed  Neumann/JS boundary conditions}

\subsec{The JS boundary conditions in terms of spins and loops}

\noindent 
The JS boundary condition can be introduced by restricting
the $O(n)$ spins on the boundary to take values in a submanifold of
`dimension' $y$.  That is, the first $y$ components the $O(\n)$ are
free, while the rest $\n-y$ components are fixed.  This is equivalent
to replacing the Boltzmann weight \actOn\ with
\eqn\actOnD{ \prod_{ <\!  rr'\!  >\in {\rm bulk} }   \ss(r) \cdot \ss(r')
\prod_{{r\in {\rm boundary} } } \ \sum_{b=1}^yS_b(r)S_b(r) .}
The Boltzmann weight \actOnD\ is invariant with respect to a subgroup
$O(y) \subset O(\n)$, which means that the JS boundary conditions are
associated with the conjugacy classes of $O(\n)$.  This is a necessary
condition to have conformal invariant boundary theory \AS.

In the original formulation of the $O(\n)$ model both $\n$ and $y$ are
integers,\foot{It is always possible to consider part of the
components of the $O(\n)$ vector as anticommuting variables, so that
the restriction $0\le y\le \n$ can be avoided.} but the result of
evaluating the trace can be analytically continued for non-integer
values of $y$ and $\n$.  The disk partition function is then
formulated as a gas of fully packed loops on the world sheet, having
two different fugacities.  The loops that do not touch the boundary
have fugacity $\n$, while the loops that are reflected from the
boundary one or several times have different fugacity $y$.

There are two classes of local operators compatible with the boundary
measure in \actOnD, which we denote by $ {\Bbb S}_L^{||}$ and $ {{\Bbb
S}}_L ^{\perp}$.  They are defined as $O(y)$ invariant polynomials of
the spin components:
\eqn\SnJS{ { {\Bbb S}}_L^{||} = \sum_{1\le a_1<\dots a_L \le
y} S_{a_1}\dots S_{a_L}, \qquad {  {\Bbb S} }_L ^{\perp}= \sum_{ y+1\le
b_1<\dots b_L \le \n} S_{b_1}\dots S_{b_L} .  }
  %

  \epsfxsize=200pt
 \vskip 10pt
 \centerline{
 \epsfbox{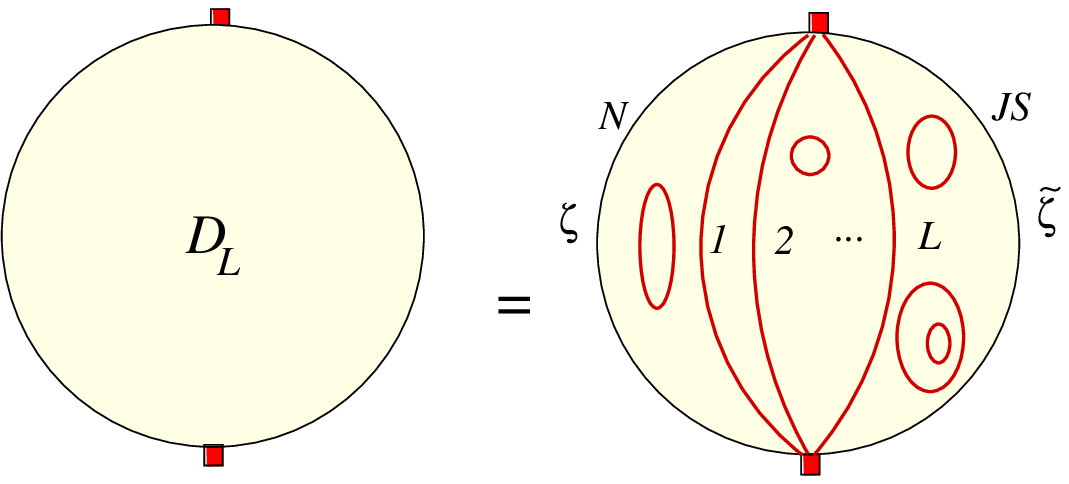    }
 }
 \vskip 5pt
 
 \centerline{\ninepoint Fig.  3: Topology of the loop configurations
 for the boundary two-point functions.  }
  
  \vskip  20pt

We will investigate the boundary correlation functions of the
operators \SnJS\ with Neumann boundary condition on the left segment
and JS boundary conditions on the right segment of the boundary.  We
denote these correlation functions by
\eqn\defDnb{D_L^{||}(\z , \tilde \z)=\<^{\z}_{_{\rm N}} [{ {\Bbb
S}}_L^{||}]^{\tilde \z} _{_{\rm JS}} [{ {\Bbb S}}_L^{||}] ^{\z}_{_{\rm
N}} \>_{\rm disk} \, , \qquad D_L^{\perp}(\z , \tilde \z)=
\<^{\z}_{_{\rm N}} [{ {\Bbb S}}_L^{\perp}] ^{\tilde \z} _{_{\rm JS}}
[{ {\Bbb S}}_L^{\perp}] ^{\z}_{_{\rm N}} \>_{\rm disk}.  }
The symbol $\< \ \ \>_{\rm disk} $ means a double sum over the $O(\n)$
spins and over the triangulations of the disk, characterized by the
boundary cosmological constants $\z$ and $\tilde \z$ associated with
the two segments of the boundary.  In terms of the loop gas, the
two-point functions \defDnb\ are the partition function of the loop
gas with $L$ open lines connecting two marked points on the boundary,
as shown in Fig.  3.  In the loop expansion of $D_L^{||}$, the
configurations where the rightmost open line touches the JS boundary
have the same weight as those in which it doesn't.  In the case of ${
S }_L ^{\perp}$, the lines that touch the JS boundary have zero
weight.  In the terminology of \JS, the correlators $D_L^{||}$ and
$D_L^{\perp}$ are the two-point functions of the $L$-leg operator
respectively in the {\it blobbed} and {\it unblobbed} sectors.

\subsec{Loop equations for the $L$-leg correlators with mixed N/JS
boundary conditions}

\noindent
The correlation functions \defDnb\ are defined as the series expansions
\eqn\DLzz{ D_L (\zeta,\tilde \zeta)= x^{-L} \sum_{l,\tilde l=0}^\infty
D ^{L} _{l,\tilde l}\ \zeta^{-l-1}\,\tilde \zeta^{-\tilde l-1}, }
where $D_L$ can be either $D_L^{||}$ or $D_L^{\perp}$.  The
coefficients $D^{L}_{ l,\tilde l} $ are the the two-point functions
with fixed boundary lengths, $l$ and $\tilde l$.  The functions $D_L$
and $D_{L-1}$ satisfy a simple recurrence relation \Kbliou\KPS. This
relation in fact holds for Neumann boundary condition on the left
segment and {\it any} boundary condition on the right segment of the
boundary.  It is obtained by taking into account all possible
configurations of the strip containing the leftmost open line (Fig.
4).

  \epsfxsize=200pt
 \vskip 20pt
 \centerline{
 \epsfbox{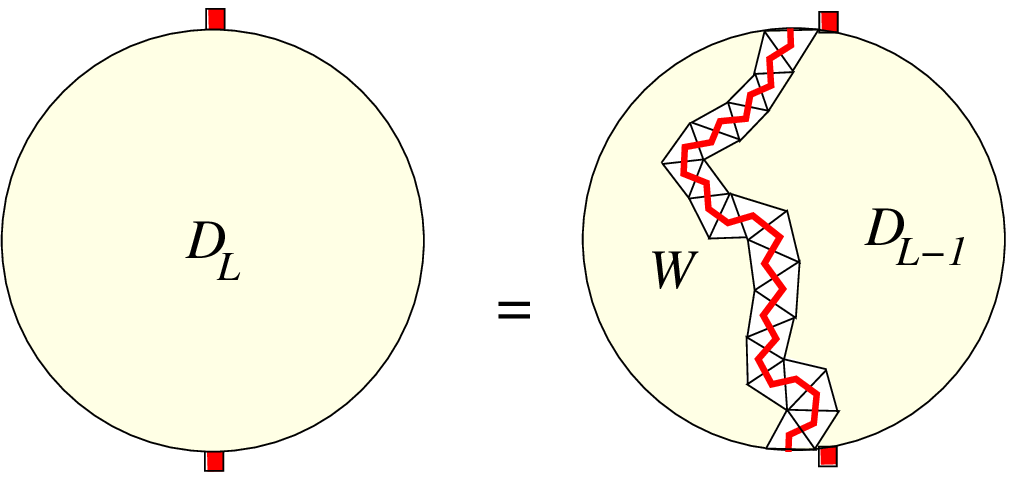    }
 }
 \vskip 5pt
 
 \centerline{\ninepoint Fig.  4: The recurrence equation for the
 boundary two-point correlators $(L\ge 2)$ }
  
  \vskip  20pt

\noindent Each such configuration is characterized by the lengths $p$
and $q$ respectively of the left and the right boundaries of the
strip.  The length of the leftmost line is then $p+q$.  The strip
splits the triangulated disk into two pieces in such a way that the
Boltzmann weights factorize, so that the sum over triangulations and
loops can be done in each piece separately.  The left piece
contributes the disk amplitude $W_{l+p}$, while the right piece yields
the boundary two-point function $D^{L-1}_{q,\tilde l}$, again with
Neumann boundary condition on the left segment.  Taking into account
that the strip in the middle can be realized in $(p+q)!/p!q!$ ways as
well as the the factor $x^{-p-q}$ associated with the area $p+q$ of
the strip we get
  \eqn\recInt{\eqalign{ & D ^{L}_{ l, \tilde l} = \sum_{p,q=0}^\infty
  W _{l+p}\ {(p+q)!\over p!q!}x^{-p-q}\ D ^{L-1}_{q, \tilde l} \, .  }
  }
Written in terms of the generating functions \DLzz, this equation
takes the form
\eqn\DLzzt{ D _{L} (\zeta,\tilde \zeta) = [W( \zeta) D _{L-1} ( x
-\z,\tilde \zeta)] _{_<}\ \, }
where we used the same notations as in \lopeqa.  We again express the
projection $[\ \ ]_<$ as a contour integral:
\eqn\recIntz{ D _{L} (\z,\tilde \z) = - \oint {d\z'\over 2\pi i} \
{W(\z) - W(\z') \over \z-\z'}\, D _{L-1}(x-\z', \tilde \z)\, .  }
This gives for the discontinuity on the real axis
\eqn\DLdisc{ {\rm Disc}_{\zeta} D _{L} (\zeta,\tilde \zeta) + {\rm
Disc}_{\zeta} W(\z)\cdot D _{L-1 } (x- \zeta,\tilde \zeta) =0.  }

Equation \recInt\ was derived for $L\ge 2$, but we will extend the
definition of the two-point functions to $L=0$, so that the it holds
also for $L=1$.  The subtle point here is how to weight the degenerate
triangulations where the left and right boundaries touch at one or
several points.  The two-point function $D_0^{||}$ is defined as a sum
over all triangulations, including those where two boundaries touch,
while in the loop expansion of the two-point function $\AA$ the
triangulations with touchings are excluded.  Such partition functions
behave as though local operators were inserted at the marked points
\CardyBB. We will think of $D_0^{||}$ and $D_0^{\perp}$ as the
boundary two-point function of the operators $S_0^{||}$ and
$S^\perp_0$, correspondingly.

The recurrence relations \recInt\ allow to determine all correlation
functions \defDnb, once the functions
$$A\equiv D_0^{\perp}  \quad { \rm and }
 \quad B\equiv D_1^{||}  $$
are known.  Below we derive two independent non-linear equations for
$A$ and $B$ and find their unique solution in the continuum limit.
These equations involve nontrivially only the cosmological constant
$\z$ of the Neumann boundary, while the cosmological constant $\tilde
\z$ of the JS boundary enters as a parameter.  Therefore from now on
the dependence on $\tilde \z$ will be implicit in our notations.  The
first equation is derived by splitting the sum of triangulations
contributing to $A_{l, \tilde l} $ into four sets associated with the
way the first edge of the JS-boundary is connected to the rest of the
triangulation (Fig.  5).  First, there is the possibility that the JS
boundary has zero length ($\tilde l=0$).  In this case $A_{l,\tilde
l}=W_l$.  If the JS boundary contains at least one edge, we count 2
possibilities.  The first edge can be glued to another edge of the JS
boundary, or it can be the edge of a triangle containing a segment of
a loop.  The last possibility is realized by two types of
configurations: (a) a loop that touches the boundary only once and (b)
a loop that touches the boundary at least twice.

  \vskip  10pt
  
   \epsfxsize=340pt
  \vskip 20pt
  \hskip 5pt
  \epsfbox{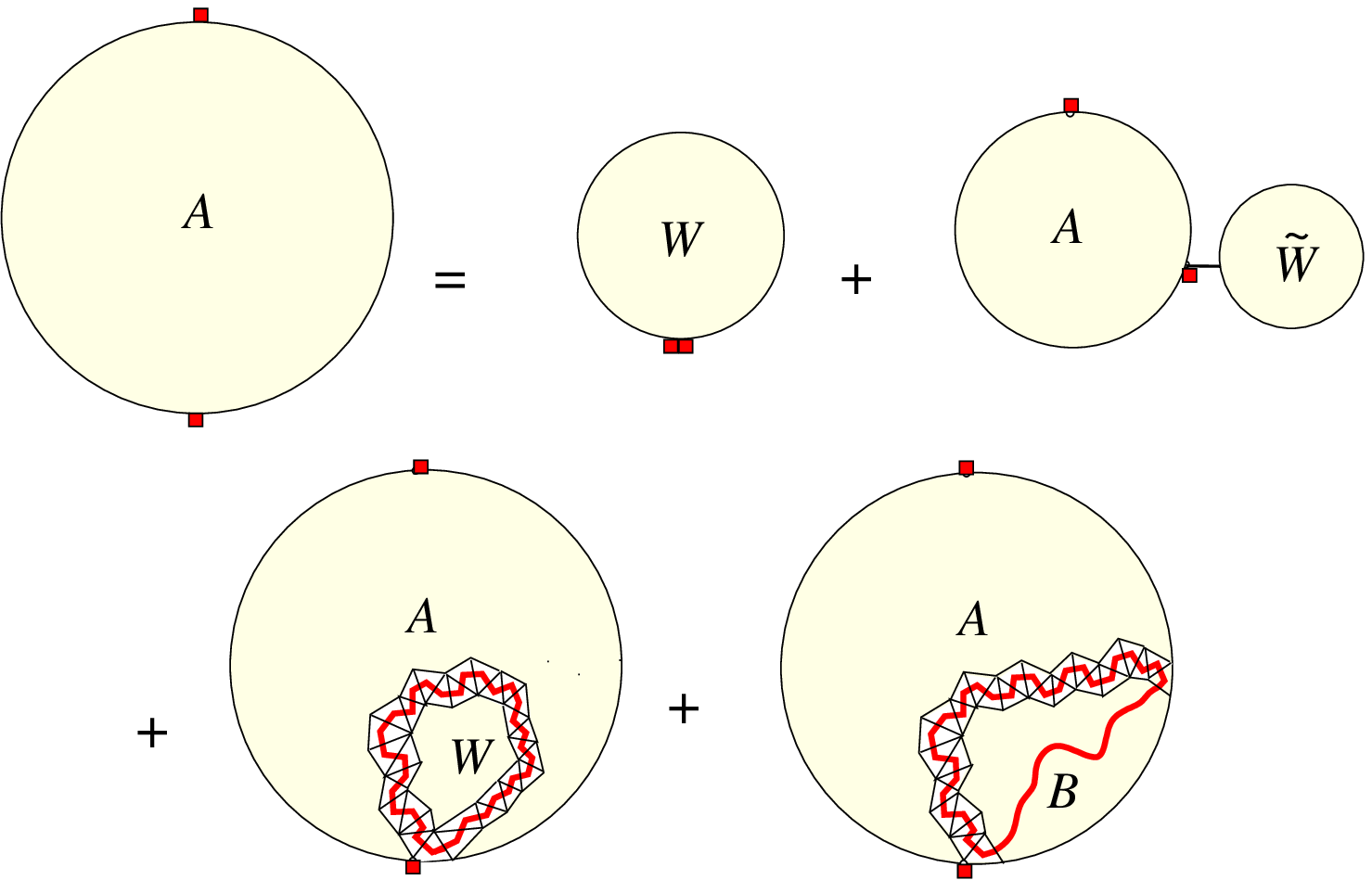    }
  \vskip 5pt
  
  \centerline{\ninepoint Fig.  5: The loop equation for $A= D_0^\perp$
  }
   
   \vskip  30pt

\noindent In the case (a) we get a product $A_{l+p, \tilde l} W_q$, as
in the loop equation \pictWw\ for the disk amplitude.  In the case (b)
we can apply the same argument as the one used in the derivation of
\recInt.  The strip that splits the disk into two here consists of the
triangles visited by the most external arc of the loop.  The left
piece is bounded by the Neumann boundary, the left boundary of the
strip of length $p$ and a piece of the JS boundary of length $k$.  It
yields a factor $A_{l+p, k}$.  The right piece is the partition
function of the disk with mixed Neumann/JS boundary conditions and an
open line connecting the extremities of the two segments of the
boundary.  The open line can touch both boundaries unrestricted number
of times.  By definition this is the two-point function $B_{q,\tilde l
-k -2}$, where $q$ is the length of the right side of the strip.

Summing up the four terms and taking into account the combinatorial
factors, we get
\eqn\leqnAB{\eqalign{ & A_{l, \tilde l}= W_l\, \d_{\tilde l, 0}+ \sum
_{k=0}^{\tilde l -2} A_{l, \tilde l - k -2} \tilde W_k\cr &+
\sum_{p,q=0}^\infty x^{ -p-q-1} \frac{(p+q)!}{ p!q!} A_{l+p, \tilde l
-1}W_{q}\cr &+ y \sum_{k=0}^{\tilde l -2} \sum_{p,q=0}^\infty x^{
-p-q-2} \frac{(p+q)!}{ p!q!} A_{l+p, k}B_{q,\tilde l -k -2}.  } }

The equation satisfied by $B$ is very similar to \recInt.  Here we
distinguish two possibilities: the open line does not touch (touches
at least once) the JS boundary (Fig.  6):
\eqn\leqnBA{ \eqalign{
 & B_{l,\tilde l} = \sum_{p,q=0}^\infty {(p+q)!\over p!q!} x^{ -p-q} W
 _{l+p} A_{q, \tilde l}\cr &+ \sum_{p,q=0}^\infty {(p+q)!\over p!q!}
 x^{ -p-q-1} B_{l+p, \tilde l - k - 1} A_{q, k} .  } }
 
 \bigskip

   \epsfxsize=330pt
  \vskip 20pt
  \hskip 5pt
  \epsfbox{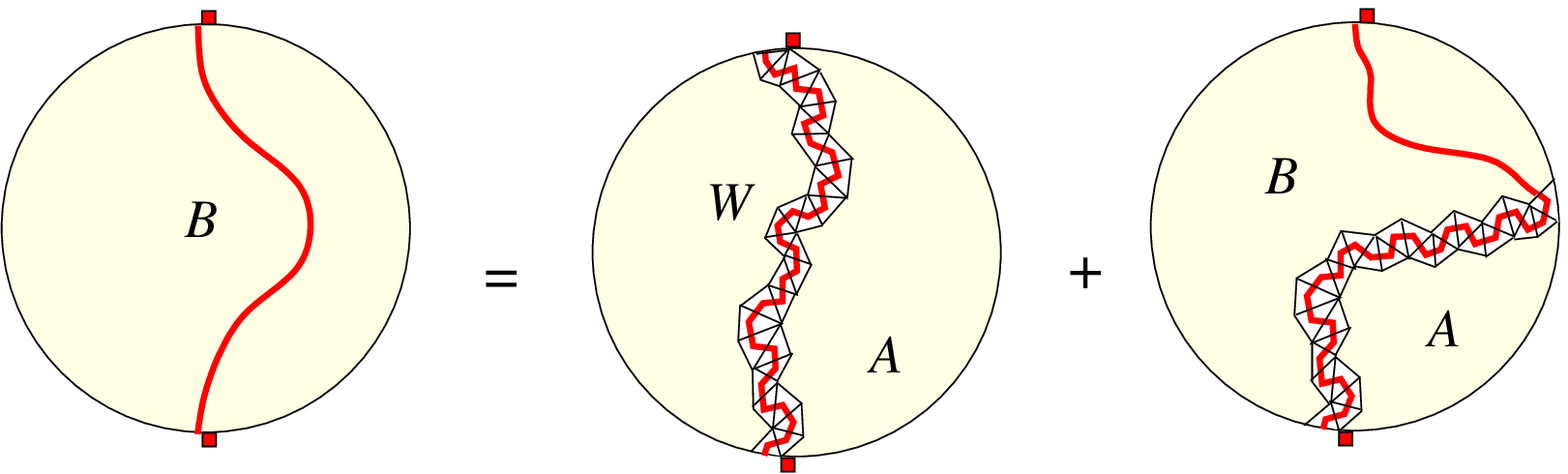    }
  \vskip 5pt
  
\centerline{\ninepoint Fig.  6: The loop equation for $B= D_1^{||}$ }
   
   \vskip  30pt

In terms of the generating functions \DLzz\ the  loop equations 
\leqnAB\ and \leqnBA\ state
\eqn\leAB{\eqalign{ \tilde \zeta A(\zeta )= {W(\zeta) }+ A(\zeta )
{\tilde W(\tilde \zeta ) } + y \[ A(\zeta ) \( W(x-\zeta) +
B(x-\zeta)\)\]_{_<} } }
\eqn\leBA{ B(\zeta)= \[ W(\zeta) A( x -\zeta) \]_< + \[B(\zeta) A( x
-\zeta)\]_< \, .}
Representing, as before, the projection $[\ \ ]_<$ as a contour
integral and taking the discontinuity across the real axis, we obtain
 \eqn\DiAB{ y {\rm Disc}_{\zeta} A(\zeta ) \cdot \( W(x-\zeta)+
 B(x-\zeta) + \frac{1}{ y} ( \tilde W( \tilde \zeta ) - \tilde
 \zeta)\) + {\rm Disc}_{\zeta}W(\zeta)=0 }
\eqn\DiBA{ {\rm Disc}_{\zeta} B(\zeta) = {\rm Disc}_{\zeta} W(\zeta)
\cdot A( {x} -\zeta) + {\rm Disc}_{\zeta} B(\zeta) \cdot A( x -\zeta).
}

\subsec{Final form of the loop equations for ${D^{||}_1} $ and $
D^{\perp}_0$}

 \noindent  The linear terms in equations \DiAB\ and \DiBA\  can be 
 eliminated by shifting  the observables $A=\AA$ and $B=\BB$.
 We redefine $\AA$ and  $\BB$ as
\eqn\newAB{\eqalign{ \BB (\z)&:= 
B(\zeta) + W(\zeta) + {1\over y} 
(\tilde W(\tilde \zeta ) - \tilde \zeta) \, ,
\cr \AA(\z)& := A( \zeta) -1
.}
 }
The terms added only change the weight when one or both boundaries
degenerate to a point and thus does not affect the critical behavior.
For the shifted quantities \newAB, equations \DiAB\ and \DiBA\
simplify to
\eqn\ABc{ {\rm Disc}_{\zeta} D^{\perp}_0(\z ) \cdot D^{||}_1( x-\z) +
{1\over y} {\rm Disc}_{\zeta} W(\z) =0\, , } \eqn\BAc{ {\rm
Disc}_{\zeta} D^{||}_1( z)\cdot D^{\perp}_0( x- \z)+ {\rm
Disc}_{\zeta} W(\z) =0.  }

The set of equations \DLdisc, \ABc\ and \BAc\ is overdetermined.
Equation \BAc\ is consistent with \DLdisc \ for $L=1$ under the
condition
 \eqn\Dpar{ D^{||}_0 = -{1\over D^{\perp}_0 }.  }
This condition has simple geometrical explanation.  Indeed, it is easy
to see that the correlation functions $D^{||}_0$ and $D^{\perp}_0$
are, unlike their counterparts on a flat lattice, different.  The
difference comes from the fact that on a dynamical world sheet the
Neumann boundary can come close to the JS boundary and touch it one or
several times.  Microscopically, the algebraic relation \Dpar\ can be
understood as a geometric progression obtained by taking into account
all possible touchings, \eqn\Dparb{ D^{||}_0= 1+ A+ A^2+\dots =
{1\over 1- A}.  }

\bigskip
   
The two equations \ABc\ and \BAc, together with the asymptotics at
infinity that follows from the expansion \DLzz, imply the following
functional identity:
\eqn\feq{ D^{||}_1(\z) D^{\perp}_0(x-\z) =- W(\z) -{1\over y} W(x-\z)
- {1\over y} ( \tilde W(\tilde \zeta ) - \tilde \zeta).  }

The loop equations \DLdisc, \ABc\ and \BAc\ obtained here allow in
principle to compute all boundary two-point correlators with mixed
Neumann/JS boundary conditions.  The three couplings, $x$, $\z$ and
$\tilde \z$, associated respectively with the area of the
triangulation, the length of the Neumann boundary and the length of
the JS boundary, enter in the loop equations implicitly through the
disk amplitude $W(\z)$.

\newsec{The continuum limit}

\noindent In this section we will study the continuum limit of the
solution, in which the three couplings are tuned close to their
critical values, $\x^*, \z^*$ and $\tilde \z^*$.  The solution in the
continuum limit depends on the three renormalized couplings $$ \mu
\sim x-x^*, \ \ \ z\sim \z - \z^* \ \ {\rm and}\ \ \tilde z \sim
\tilde \z-\tilde \z^*.
 $$

\subsec{Disk one-point function  with Neumann boundary conditions}

\noindent
The power series  \lpgz\  converges for $x>x^*$, where 
$$x^* = 2\sqrt{2(2+n)}
$$
is the the critical value of the cosmological constant \DTS. We
introduce a small cutoff parameter (elementary length) $a$ and define
the renormalized cosmological constant $\mu$, boundary cosmological
constant $z$ and loop amplitude $w$ as follows\foot{ Note that in the
dense phase the boundary has anomalous dimension.  This is a
consequence of the fractal structure of the boundary in this phase.}
\eqn\vvvt{
\eqalign{
 \mu:& =  a^{-2}\,  {8 \over 4-\n^2}
 \,  {x^2-{x^*}^2 \over  {x^*}^2}\, , \cr
 z: &=  a^{-1/ (1-\th)}   \(\z - \hf x\)  ,\cr
  w(z) :&= a^{- 1}  \(W(\z)- {1\over 2-n} \z +
{n\over 4-n^2} \, x\) . }
}
Then equation \wwt\ takes the form
\eqn\lesl{ w(z)^2+ w(-z)^2 + n w(z)w(-z) = \mu \, \sin^2\pi\th\, +
a^{2\th/(1-\th)}\, z^2/( 2-n)\, .  } 
In the continuum limit $a\to 0$, the second term on the r.h.s. of
\lesl\ can be neglected provided $\half < \th<1$, or $0<\n<2$.  Then
the solution of \lesl\ can is written in parametric form as
\eqn\xoft{\eqalign{ z\ \ \ =&\ \ \ M\, \cosh\t\, , \cr w(z ) =& - \,
M^{1-\th} \, \cosh (1-\theta )\t \, ,} }
where $ M= C \mu ^{1/2(1-\th)}$.  The value of the constant $C$ is
fixed by the normalization of the disk partition function $\Phi$.
One possible choice is  
\eqn\solutionW{ \p_\mu \Phi = {M^{\th}\over \th} \cosh\th \t, \quad
\p_z \Phi = - w(z), \quad M^{2-2\th} = 2\mu .  } 
As a function of $z$, the loop amplitude $w(z)$ has a branch cut along
the interval $[-\infty, -M]$.  The solution \xoft\ corresponds to
Liouville gravity with matter central charge
\eqn\cden{ c=1-6 {\theta ^2\over 1-\theta } \, .}
The susceptibility $u(\mu)$, which is by definition the partition
function of the loop gas on a dynamically triangulated sphere with two
punctures, scales as $u\sim \mu^{-\gst}$, with
\eqn\gamastr{\gst= -{\theta \over 1-\theta }.  }

\subsec{Boundary  two-point functions of the $L$-leg operators}

\noindent\ The continuum limit of the boundary two-point functions is
obtained as a triple scaling limit in $x, \z$ and $\tilde \z$, in
which the area of the triangulation as well as the lengths of the
Neumann and JS boundaries diverge.  The point $x = x^*, \z=\z^*,\tilde
\z = \tilde\z^*$ is a singular point of equation \feq, where the
r.h.s. vanishes: \eqn\crittz{ W(\z^*) +{1\over y} W(\z^*) +{1\over y}
( \tilde W(\tilde \zeta ^*) - \tilde \zeta^*)=0\, .  } We define the
renormalized cosmological constant $\tz$ and loop amplitude with JS
boundary conditions as
\eqn\deflb{\eqalign{
\tilde z:&=  a^{-1/ (1-\th)}   \( 
\tilde \zeta-\tilde \zeta^*\)\, , \cr
\quad \tilde
w(\tilde z) :&= a^{-1}\( \tilde W(\tilde \z)- \tilde W(\tilde \z^*) \).  }
}
In the continuum limit $a\to 0$, the linear terms in $z$ and $\tilde
z$ can be neglected, since they are multiplied by $a^{\th/(1-\th)}$,
and functional equation \feq\ takes the form
 \eqn\loopc{ D^{||}_1(z) D^{\perp}_0(-z) =- w(z) - {1\over y} w(-z) -
 {1\over y} \tilde w(\tilde z) .  }
 Together with \Dpar, this equation yields a linear relation between
 $D_1^{||}$ and $D_0^{||}$:
 \eqn\Dparc{ D^{||}_1(z) =\[ w(z) + {1\over y} w(-z) + {1\over y}
 \tilde w(\tilde z) \] D^{||}_0(-z) .  } 
 Equation \Dparc\ is the key result of this paper.  It allows to
 determine, up to a normalization, the boundary two-point functions
 with $ D^{||}_1$ and $ D^{\perp}_0.  $ The two-point functions with
 $L>1$ are then easily evaluated from the recurrence equations
 \DLdisc, which have, in the continuum limit, the form
\eqn\reccont{ D_L(z+i0) -D_L(z-i0)= [w(z+i0) -w(z-i0)] \, D_{L-1}(z)
\qquad (z<-M) \, .  }
The solution, as a function of $\mu$,
  $z$ and $\tz$, is
expected to be of the form
\eqn\genLc{ D_L(\mu, z,\tz)=\mu^{1- {1\over 2} \gst} (\sqrt{\mu}\,
)^{2\DB_L-2} \ \hat D_L\( {z/ M}, {\tilde z/ M}\) , } 
where the exponent $\gst$ is given by \gamastr, $\DB_L$ is the
boundary gravitational dimension of the $L$-leg operator, and $M =
(2\mu)^{1/(2-2\th)}$.

Before giving the complete solution, we are going to solve a simpler
problem: to determine the scaling behavior of the two-point function
when $\mu=\tilde z=0$.  Restricted in this way, the two-point function
\genLc\ reduces to a power of the only non-zero coupling $z$:
\eqn\DLz{ D_L
  \sim z ^{(1-\th)(2\DB_L- \gst)} }
From here one can determine the conformal dimensions of the $L$-leg
operators using the KPZ map \KPZsc .

 \subsec{Evaluation of the $L$-leg critical exponents for N-JS
 boundary conditions}

\noindent At $\mu=\tz=0$, the solution for the 
observables   $D_L$ and $w$  must be given by powers of $z$,
\eqn\singD{ D_L^\perp \sim z^{\a_{L}}, \quad D_L^{||} \sim
z^{\b_{L}},\quad w\sim \, z^{1-\th}\, .  }
Then \loopc\ yields three identities relating $\b_0, \b_1$ and $y$.
The first one follows from comparing the powers of $z$ on both sides:
\eqn\abo{ \b_1+\a_0=1-\th\, .  }
The other two identities arise when equating the imaginary part of
both sides of \loopc\ respectively for $z>0$ and $z<0$:
\eqn\eqexp{ \sin(\pi \a_0 ) =- {1 \over y} \sin\th \pi ,\quad
\sin(\pi \b_1 ) =- \sin(\th\pi ).  }
These equations determine $y$ as a function of $\a_0$:
\eqn\mab{ y = {\sin \pi \b_1 \over\sin \pi\a_0 } = {\sin \pi ( \a_0+
\th ) \over\sin \pi\a_0 } .  }
We see that the expression \mab\ for $y$ is of the form \defy\
with
\eqn\alfr{ \a_0 = r\th\, .  }

To determine $\a_0$, we must invert the multi-value function $y(\a_0)=
y(\a_0+1)$.  The relevant branch is given by the lowest positive value
of $\a_0$, which is must be in the interval \eqn\aoineq{0\le \a_0 < 1
\qquad {\rm or} \quad 0<r<{1\over \th } \ .  } The fact that $\a_0$ is
positive is a consequence of the representation \Dparb\ of $ D_0^{||}
$.  The singularity of the l.h.s. cannot not be weaker than the
singularity of each of the terms, hence $\b_0\le \a_0$.  Since
$\a_0+\b_0=0$, this means that $\a_0$ is positive and $\b_0$ is
negative.\foot{Let us stress that this argument is justified only for
non-negative couplings, when all terms in the series are
non-negative.}
 
Once we determined the critical exponent for $L=0$, 
the other  exponents    follow
from the recurrence relation  \reccont:
\eqn\defar{\eqalign{ \a _{L} &= L(1-\th) + r\th \qquad \b_{L}=
L(1-\th) - r\th \, .  } }
Comparing these values with \DLz\ we find for the gravitational
dimensions $\D_L^{||}$ and $ \D_L^{\perp} $
\eqn\grdimL{\eqalign{ \D_L^{\perp}= {\a_L - \th\over 2(1-\th)} &=
{L(1-\th) + r\th - \th\over 2(1-\th)} =\Delta_{r, r-L} \, , \cr
\D_L^{||} ={\b_L - \th \over 2(1-\th)} &= {L(1-\th) - r\th - \th\over
2(1-\th)} =\Delta_{-r, -r-L}\, .  } }
By the KPZ scaling relation \KPZsc, and using the symmetry
$h_{-r,-s}=h_{rs}$, we find the conformal weights of the $L$-leg
operators with mixed Neumann-JS boundary conditions:
\eqn\hdenph{ h_{L}^{||}= h_{r,r+L}, \qquad h_L^{\perp} = h_{r, r-L} ,
}
in accord with \JS .
           
When $r=1$, then $y=\n$ and the loops touching the boundary have the
same fugacity as the loops in the bulk.  In this case one obtains the
well known $L$-leg exponents with Neumann-Neumann boundary condition
on flat \DupSal\ and dynamical \KKopen \ lattices.
  $$
  h_{_{L\, {\rm  leg}}} ^{N/D}=  h_{0, 1/2+L}
  $$

\subsec{Complete solution  and relation to Boundary Liouville theory}

\noindent In the previous subsection we determined the prefactor in
\genLc.  Now we will evaluate  the scaling functions $\hat D_L$.  First,
using the expression  \xoft\ of the loop amplitude $w(z)$ and the
identity
\eqn\wwy{ \eqalign{ [w(\t ) + y^{-1} w(\t+i\pi) ]/ M^{1-\th}
  &=- C\, \cosh[(1-\th)\t + i \pi r\th] \, , \cr
 }
 }
with $C= { \sin\pi\th / \sin(r+1)\pi\th}$, we write \Dparc\ as
\eqn\fuctau{ \hat D^{||}_1(\t  ) = - \(C\, \cosh\[(1- \th)\t
+ i \pi r\th\] + \, {\tilde w(\tilde z)\over y M^{1-\th}}\) \, \hat
D^{||}_0(\tau+i\pi) \, .  }
Next,  we change the variable $\tilde z\to \tilde \t$ so that
\eqn\twtztt{
    {\tilde w(\tilde z)\over M^{1-\th}}
    = y \, C\, \cosh (1-\th)\tilde\t\, 
    =  {\sin \pi\th \over \sin \pi r \th}   \, \cosh (1-\th)\tilde\t\ .
    }
 After that equation  \reccont\ takes the form
 \eqn\fuctaut{\eqalign{ \hat D^{||}_1(\t ,\tilde \t ) &= -C\, \[
 \cosh\((1- \th)\t + i \pi r\th\) + \cosh(1-\th) \tilde \t)\] \, \hat
 D^{||}_0(\tau+i\pi, \tilde \t) 
 \, .  } }
Let us remark that equation \twtztt\ is just a change of variable and
does not involve any assumption about the the boundary one-point
function with JS boundary conditions $\tilde w(\tilde z)$, because the
function $\tilde z (\tilde \tau)$ is not yet determined.  On the other
hand, the only solution compatible with the world-sheet CFT is $\tilde
z \sim M\cosh \tilde\t$.  Indeed, there is only one non-zero boundary
one-point function, that of the Liouville-dressed identity operator,
which is given by \xoft.

For generic $r$, one can recognize in \fuctaut\ the functional
equation derived by V. Fateev, A. Zamolodchikov and Al.  Zamolodchikov
for the boundary two-point function in boundary Liouville theory \FZZb
.  This equation also appeared as the first member of an infinite
series of functional identities characterizing the boundary ground
ring in 2D quantum gravity \bershkut\BGR.

The boundary two-point function in Liouville gravity depends on the
the target-space momentum $p$ and the two boundary parameters $\t$ and
$\tilde\t$.\foot{Our notations are related to the notations of \FZZb\
by $ 1/b+b-2\b = p/b,\ \t= \pi s/b, \ b^2 = 1-\th .$} It is given, up
to a normalization factor that depends only on $p$, by \FZZb
   \eqn\dbdefi{ \eqalign{ D (p; \t, \tilde \t) &= \mu^{p/2} \ \hat D
   (p; \t, \tilde \t) ,\cr \hat D (p; \t, \tilde \t)&= \exp\left( -{
   \int\limits_{-\infty}^{\infty}} {dt\over t} \left[ { \sinh (\pi p t
   /b^2) \ \cos (\t t) \cos ( \tilde \t t)\over \sinh (\pi t)\ \sinh
   (\pi t/ b^2) } -{p \over\pi t}\right] \right)\, .  } }
 The scaling function $\hat D (p; \t, \tilde \t)$ satisfies the
 identity
\eqn\FZZfed{ \hat D(p+b^2;\t,\tilde \t) = \hf \[ \cosh\left( b^2 \tau \mp i\pi
{p} \right)+ \cosh\left( b^2\tilde \tau \right)\]\hat D(p;\t\pm i\pi
,\tilde\t)\, , } 
which is the same as  \fuctaut, with $b^2 = 1-\th$,
 $p= -
r \th $ and  $C=1/2$.   Therefore 
  \eqn\Dppo{ \hat D^{||}_0(\t,\tilde\t)  = \hat D(-\th r;\t,\tilde\t), 
  \quad \hat D_1^{||}(\t,\tilde\t)  =
\hat D(-\th r +1-\th;\t,\tilde\t) }
is a solution of  \fuctaut.   The relation  \Dpar\ then implies
\eqn\Doperp{
\hat D^{||}_0(\t,\tilde\t)  = \hat D(\th r;\t,\tilde\t)\, .
}

Is this the physical solution that corresponds to the series expansion
\DLzz ?  Equation \FZZfed\ does not determine uniquely the two-point
function.  In order to specify the unique solution, FZZ \FZZb\ used
the duality symmetry, which supplies another equation of the same
type, as well as the symmetry in $\t\leftrightarrow \tilde\t$.
Neither of these symmetries is satisfied in the microscopic theory.
If we assume that the physical solution enjoys these symmetries in the
scaling limit, then \Dppo\ is the unique solution.  In the particular
cases $r=1$ and $r= (1-\th)/2\th$, the function \Doperp\ reproduces
correctly the expressions obtained previously for the disk partition
function with mixed Neumann/Neumann and Neumann/Dirichlet boundary
conditions.

Once the two-point functions with $L=0$ are known, the rest can be
determined from the recurrence equation \reccont, which can be cast
into the form
\eqn\fdeqV{\eqalign{ \hat D _{L} (\t+i\pi , \tilde \t)-\hat D _L
(\t-i\pi , \tilde \t) &=- 2 i \sin\pi\th \, \sinh[(1-\th )\t]\, \hat D
_{L-1} (\t, \tilde \t). } }
This equation  is compatible (up to normalization) with \FZZfed\ upon the
identification
\eqn\pparper{ p^{||}_L= -\th r + (1-\th)L  \quad  {\rm and}
\quad p_L^{\perp}= \th r + (1-\th) L
\, , }
 respectively for $D^{||}_L$ and $D^{\perp}_L$.  When $L\ge 1$, the
 operators ${\Bbb S}_L^{||}$ and ${\Bbb S}_L^{\perp}$ have different
 conformal weights.  The two-point functions $D^{||}_0$ and
 $D^{\perp}_0$ are related by Liouville reflection, and correspond to
 the `physical' and `unphysical' Liouville dress of the same boundary
 matter field with conformal weight $h_{r,r}$.

   \subsec{Dirichlet boundary conditions and twist operators}

 \noindent The Dirichlet boundary conditions for the $O(n)$ model is
 defined by fixing the $O(\n)$ spin on the boundary to point to given
 direction, say $\ss=(1,0,\ldots 0)$.  This is a particular case of
 the JS boundary condition, obtained by taking $y=1$, or equivalently
\eqn\rtwist{ r={1-\th\over 2\th} \, .  }
This case deserves special attention, because here we can compare the
general formula \Doperp \ with the exact results found in
\KKopen\Kbliou\KPS.

            \vskip 0.9cm

   \epsfxsize=150pt
  \hskip 85pt
  \epsfbox{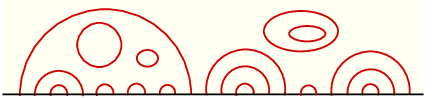    }
  \vskip 15pt
  
\centerline{\ninepoint Fig.  7: A loop gas  configuration  for 
Dirichlet boundary condition.  }
   
   \vskip  10pt

 The Dirichlet boundary condition for the $O(n)$ model leads to the
 same loop gas expansion as the Neumann boundary condition for the SOS
 model, which was first studied in \KKopen\ and then given a world
 sheet CFT interpretation in \Kbliou\ and \KPS. Namely, it is assumed
 that each point of the boundary is an endpoint of an open line, as
 sketched in Fig.  7.  The boundary condition changing operator $
 {\Bbb T}$ separating Dirichlet and Neumann boundary conditions was
 called in \Kbliou\ {\it twist operator}, by analogy with the gaussian
 field.  The correlation function of two twist operators,
  \eqn\corfT{ 
  \Omega(z,\tilde z)=  \<  _{\rm N} ^{z} [ {\Bbb T}]^{\tilde z}_{\rm D} 
  [ {\Bbb T}] _{\rm N} ^{z} \>_{\rm disk},
    } 
  was first evaluated in \KKopen, eqs.(4.34)-(4.37) of that paper.
  Afterwards this solution was identified \Kbliou\ as special case of
  the FZZ two-point function \dbdefi\ with $p = (1-\th)/2$, which
  corresponds to $r$ given by \rtwist.

  It is easy to see that the sum over the configurations with open
  lines as the one in Fig.  7 can be interpreted, for this particular
  value of $r$, either as the loop expansion for $\AA$, or as the loop
  expansion for $\BB$.  Indeed, if we connect pairwise the endpoints
  of the open lines, as shown in Fig.  8a, we obtain a configuration
  of the loop expansion for $\AA$.  The Boltzmann weights also match
  under the condition that all loops that touch the boundary have
  fugacity $y=1$.  Alternatively, we can leave the first and the last
  open line endpoint free and connect the rest of the endpoints
  pairwise, as is shown in Fig.  8b.  Then the first and the last
  points are connected by an open line, and we obtain a configuration
  of the loop expansion for $\BB$.  Therefore, even at microscopic
  level,
$$\Omega=\AA = \BB\, \qquad (y=1).$$

   \epsfxsize=150pt
  \vskip 20pt
  \centerline{\epsfbox{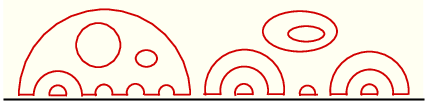 } \hskip 1cm \epsfxsize=150pt
  \epsfbox{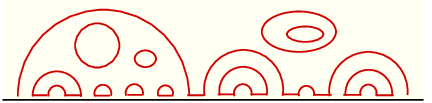 }
  }
  \centerline{\hskip 3cm a \hskip 6cm b \hskip 4cm}
  \vskip 5pt
  
\centerline{\ninepoint Fig.  8: The two ways of closing the open
lines.  }
   
   \vskip  10pt

The two-point function of the boundary twist operator, $\Omega(\t,
\tilde \t)$, is obtained as the solution of a quadratic functional
identity, eq.  (4.25) of \KKopen, which is identical to \loopc\ with
$y=1$.  In this particular case the solution can be found without
additional assumptions about symmetry, and the result \KKopen\
coincides with \dbdefi\ for $p = (1-\th)/2$.
   
The conformal weights of the excited twist operators $ {\Bbb T}_L$, or
the boundary $L$-leg operators with mixed Neumann/Dirichlet boundary
conditions, were identified in \KPS\ as\foot{In \KPS, the conformal
weights for the excited twist operators were actually evaluated for
the dilute phase, where they are given by $h_{L+1/2, 0} $.  }

\eqn\dimTL{  h_{{T}_L} = h_{0,L+{1/2} }\,  .     } 
On can  check that 
$  h_{{T}_L}=  h^{||}_{L+1} = h^{\perp}_L\, , 
$
with $r$ given by \rtwist. The different Kac-table like  identifications 
of these operators are possible because of the ambiguities of the 
representation \scaldims.

Thus the results obtained here for general $r$ are in full agreement
with those of \refs{\KKopen, \Kbliou, \KPS}.  There is however a
difference in the interpretation of the results, which is due to the
different form of the microscopic loop equations.  Compared to
equation (4.18) of \KKopen, our equation \feq\ contains an extra term
$\tilde W$, which takes into account the self-touchings of the JS
boundary.  This term was omitted in \KKopen, hence the self-touchings
of the Dirichlet boundary were not taken into account there.  As a
result, the two-point function \corfT, evaluated in \KKopen, describes
a sum over surfaces with self-touchings allowed for the Dirichlet
boundary and forbidden for the Neumann boundary.  This explains the
puzzling observation, made in \KKopen, that the fractal dimensions of
the Dirichlet and Neumann boundaries are different for $\th \ne 0$.
 
We find here more natural to define the sum over surfaces so that both
segments of the boundary have the same dimension, which is the case
when the contact term in question is taken into account.

\newsec{Conclusions}

\noindent In this paper we evaluated the boundary two-point function
for the $O(\n)$ loop model on the dynamically triangulated disc with
presumably the most general boundary conditions, constructed recently
in \JS. We restricted ourselves to the dense phase of the loop gas,
where both the bulk and the boundary are critical and the only
parameters of the theory are the bulk and boundary cosmological
constants.  The scaling behavior of the two-point function confirms
the $L$-leg exponents \hdenph\ conjectured in \JS. Our result for the
two-point function implies the symmetry
 \eqn\sym{ { {\Bbb S}}_L^{||} \ \leftrightarrow\ { {\Bbb S} }_L
 ^{\perp},\quad y \ \leftrightarrow\ \n-y.  }
which looks quite natural given the definition \SnJS\ of these
operators and resembles the duality symmetry that exchanges Dirichlet
and Neumann boundary conditions.  In the parametrization \defy,
exchanging $ y $ and $\n-y$ is equivalent to changing the sign of $r$,
due to the identity
\eqn\rminusr{ y (r) + y(-r) = \n.  }
In the loop gas formulation, the symmetry \sym\ is spelled out as
\eqn\blun{ \{ {\rm blobbed}, r\} \leftrightarrow \{ {\rm unblobbed},
-r\} } 
and is respected by the exponents \starde.

The dilute phase is more intricate because for each $y$ there is a
one-parameter family of boundary conditions and a fine tuning of the
matter coupling constants should be done bth in the bulk and on the
boundary.  Our preliminary results for the scaling dimensions \KoZa\
seem to be compatible with the unpublished results of Jacobsen and
Saleur.

Finally, let us mention that the $O(\n)$ model coupled to 2D gravity
can be viewed as solvable model of bosonic string theory with curved
target space, representing the $(n-1)$-dimensional sphere.  The
continuous spectrum of D-branes in this theory is presumably related
to the fact that the target space curvature,
 $$R= (n-1)(n-2),$$
is negative in the interval $1<n<2$.
 We believe that all our results can be also
obtained also on the basis of the dual $O(n)$ invariant matrix model \Ion.

\bigskip

{\bf Acknowledgements}

 {\ninepoint The author thanks J. Jacobsen and H. Saleur for
 discussions and explaining me some of their unpublished work on this
 subject.  Part of this work was carried out at the Kavli Institute
 for Theoretical Physics, Santa Barbara, supported in part by the
 National Science Foundation under Grant No.  PHY99-07949, in the
 course of the program on Stochastic Geometry and Field Theory: From
 Growth Phenomena to Disordered Systems.  This work has been partially
 supported by the European Network ENRAGE (contract
 MRTN-CT-2004-005616) and the ANR program GIMP (contract
 ANR-05-BLAN-0029-01).  }

 \listrefs
 \bye